\documentclass[journal]{IEEEtran}
\usepackage{amsmath,amssymb,amsfonts}
\usepackage{graphicx}

    \usepackage[left = 1.95 cm, right = 1.95 cm, top = 1.95 cm, bottom = 1.95 cm]{geometry}

\usepackage[english]{babel}
\usepackage[utf8]{inputenc}
\usepackage{algorithm}
\usepackage[noend]{algpseudocode}

 \usepackage{cite}

\ifCLASSINFOpdf
\else
\fi
\usepackage{amsmath,amssymb,amsfonts}

\usepackage{graphicx}
\graphicspath{%
    {converted_graphics/}% inserted by PCTeX
    {/}% inserted by PCTeX
}
  \makeatletter
\newcommand*{\rom}[1]{\expandafter\@slowromancap\romannumeral #1@}
\makeatother

\usepackage{xcolor}
  
\begin{document}
  \bstctlcite{IEEEexample:BSTcontrol}

\title{Optimizing Cognitive Networks: Reinforcement Learning Meets Energy Harvesting over Cascaded Channels
}

  \author{Deemah H. Tashman,~\IEEEmembership{Member, IEEE}, Soumaya Cherkaoui,~\IEEEmembership{Senior Member, IEEE}, and Walaa Hamouda,~\IEEEmembership{Senior Member, IEEE}
\thanks{D. Tashman and S. Cherkaoui are with the Department of Computer  and Software Engineering, Polytechnique Montreal, Montreal, QC, Canada,  H3T 1J4 (e-mail: \{deemah.tashman, soumaya.cherkaoui\}@polymtl.ca).} 
\thanks{ W. Hamouda is with the Department of Electrical and Computer Engineering, Concordia University, Montreal,  QC, Canada, H3G 1M8, (e-mail:  hamouda@ece.concordia.ca).  }}

 \maketitle

\begin{abstract}
This paper presents a reinforcement learning (RL) based approach to improve the physical layer security (PLS) of an underlay cognitive radio network (CRN) over cascaded channels. These channels are utilized in highly mobile networks such as cognitive vehicular networks (CVN). In addition, an eavesdropper aims to intercept the communications between secondary users (SUs). The SU receiver has full-duplex and energy harvesting capabilities to generate jamming signals to confound the eavesdropper and enhance security. Moreover, the SU transmitter extracts energy from ambient radio frequency signals in order to power subsequent transmissions to its intended receiver. To optimize the privacy and reliability of the SUs in a CVN, a deep Q-network (DQN) strategy is utilized where multiple DQN agents are required such that an agent is assigned at each SU transmitter. The objective for the SUs is to determine the optimal transmission power and decide whether to collect energy or transmit messages during each time period in order to maximize their secrecy rate. Thereafter, we propose a DQN approach to maximize the throughput of the SUs while respecting the interference threshold acceptable at the receiver of the primary user. According to our findings, our strategy outperforms two other baseline strategies in terms of security and reliability.

\end{abstract}


\IEEEpeerreviewmaketitle

\section{Introduction}
\par\IEEEPARstart{A}{} cognitive radio network (CRN) is a wireless communication network that effectively utilizes the available spectrum by dynamically adapting its transmission parameters and obtaining access to frequency bands \cite{tashman2023security}. CRNs are intended to address the problem of spectrum scarcity by allowing secondary users (SUs) to access the spectrum bands licensed to primary users (PUs) without causing interference \cite{10466378}. CRNs rely on spectrum sensing, spectrum management, and spectrum sharing to enable SUs to detect and utilize frequency bands without interfering with licensed users \cite{9237455}. Moreover, access requirements for SUs must include interference management, spectrum sensing, and power control to ensure efficient coexistence with PUs.  Multiple access modes are utilized by SUs in CRNs to maximize spectrum utilization while minimizing interference with PUs. For instance, underlay access mode permits SUs to transmit at reduced power levels within PU bands. Overlay paradigm permits SUs to use frequency bands concurrently with PUs in exchange for assisting the PUs in their transmissions. Additionally, SUs in interweave mode can only access  frequency bands detected by spectrum sensing as long as they are vacant. Indeed, SUs in underlay mode must continually adjust their transmission power to avoid interfering with PUs. This varies the reliability of the SUs' channels in underlay CRNs, exposing them to a variety of physical layer  threats, such as eavesdropping \cite{8904352}, rendering safeguarding SUs in CRNs  crucial.

Physical layer security (PLS) provides an additional layer of protection against unauthorized access by ensuring that the connection between legitimate ends (main link) is more reliable than that between the transmitter and an eavesdropper (wiretap link). Multiple methods, including cooperative relaying and cooperative jamming, can improve PLS \cite{9612017}.  These processes require energy conduction, which is challenging for energy-constrained devices in particular \cite{9500621,9838746,10368012,10278964,10464644}. Energy harvesting (EH) techniques resolve the power limitations of cognitive radio devices, including cognitive vehicular networks (CVNs) by converting ambient energy sources into usable electrical power \cite{9926102}. By incorporating EH mechanisms, cognitive radio devices can become self-sufficient, extend their operational longevity, and reduce their reliance on external power sources, thereby improving network efficiency and resilience. These technologies collectively improve the security, reliability, and lifespan of CRNs. Among the available  EH techniques is simultaneous wireless information and power transfer (SWIPT), which is based on the premise that radio frequency (RF) signals contain both information and energy. To render the SWIPT strategy feasible, the receiver needs to employ an EH receiving scheme, such as power splitting (PS). Therefore, a receiver employing the PS strategy stores a portion of the received energy for future use, while the remainder is employed for information decoding.

CVNs have been suggested to provide a reliable method for addressing certain obstacles for vehicle users \cite{6985745}. That is,  the demand for vehicle communication is growing exponentially, and the allocated frequency band may not be sufficient to meet the needs of vehicle users. Hence, CVNs have been designated to tackle this problem by allowing the nodes of these networks to access the licensed band.
In CVNs, all nodes are mobile and presumably reside in scattered areas \cite{8820071}. Prior research presumed that the channels in CVNs were modeled as conventional fading channels, such as the Rayleigh fading model. Nevertheless, experimental investigations indicate that cascaded channels are more suitable than non-cascaded channels for modeling these types of networks \cite{8820071}. Classical channels have proven incapable of modeling the interconnections for mobile devices \cite{4776427}. Consequently, cascaded fading channels are more accurate at modeling signals transmission \cite{9237455,9385753,ghareeb2020statistical}.

Several machine learning approaches have been employed recently to optimize CRNs. In particular, reinforcement learning (RL) has been deployed as an effective approach to improve the  reliability of the underlay architecture of CRNs. RL enables SUs to adapt their transmission strategies in response to fluctuating network conditions and potential security threats by leveraging its ability to learn from interactions with the environment \cite{9318243,9729992,9524882,10078092,9685236,9839316,9999295}. SUs can learn optimal power control policies, channel selection strategies, and access protocols to optimize their communication efficiency while minimizing their exposure to eavesdropping and interference \cite{lin2023deep}. Furthermore, RL enables SUs to modify their behavior dynamically based on observed rewards or penalties, enabling them to operate robustly and securely.

\subsection{Related Works}
  Recently, RL has been employed to improve the PLS for multiple wireless networks.  For instance, in \cite{9964272}, the authors examined mmWave communications enabled by unmanned aerial vehicles (UAVs) from a PLS standpoint. The beamforming vector and UAV trajectory, in addition to user scheduling, were concurrently optimized with statistical channel state information (CSI) of eavesdroppers to minimize the weighted sum of UAV flight time and secrecy outage duration. A method based on deep reinforcement learning (DRL) was used to optimize all variables simultaneously. In \cite{9360739}, the authors investigated the security of a UAV-assisted selective relaying wireless network in which multiple UAVs serve as decode-and-forward (DF) relays connecting a ground base station with multiple legitimate users on the ground in the presence of a passive eavesdropper. The Q-learning method was utilized to reduce the outage probability. Moreover, the work in \cite{9473714} examined UAV-to-vehicle communications susceptible to multiple ground eavesdroppers in urban scenarios. The primary objective was to maximize secrecy rates while taking into account energy consumption and flight zone limitations. Subsequently, a curiosity-driven DRL algorithm was used to address the issue. In \cite{kamboj2021intelligent}, a machine learning algorithm for relay selection in order to enhance the PLS of a dual-hop non-regenerative wireless cooperative network was proposed. In order to maximize the secrecy rate, this paper proposed two Q-learning techniques for different scenarios. Recently,  the authors in \cite{lin2023deep}  investigated the threat to confidential communications posed by a single eavesdropper in  CRNs-based EH, where both the SU and the jammer harvest, store, and use RF energy from the PU transmitter. To maximize the long-term performance of secret communication, the primary objective was to optimize the time periods for EH and wireless communication for both the secure user and the jammer. The authors proposed a multi-agent DRL method for optimizing resource allocation and performance.

\subsection{Motivation \& Contribution}
Although it has been demonstrated that RL techniques have a significant impact on boosting the reliability of wireless systems, they have not been extensively discussed for enhancing the PLS for wireless networks, particularly for CRNs. In particular,  in contrast to \cite{lin2023deep}, we presume that the SU transmitter harvests energy from ambient RF sources and not from the PUs' transmissions, which implies that the SU battery level is unaffected by the status of the PUs' transmissions.  Moreover, in our work, we assume that the SU destination is assisting in enhancing security rather than an external SU. This has several advantages, including the fact that the SU receiver can distinguish between its messages and the jamming signals and, as a result, cannot interfere with the SUs' communication. Another advantage is that this method reduces the system's complexity and saves resources that would otherwise be required to operate an external jamming SU.
%As a security metric, the secrecy capacity is utilized, whereas the throughput is employed to evaluate the system's reliability. Additionally, we compare our work to other benchmarks to demonstrate the efficiency of our model.

No work, as far as the authors are aware, has considered the deep Q-network (DQN) approach to enhance the PLS for an underlay CVN with  EH capabilities over cascaded channels, which motivated us to address this  gap in this paper. Therefore, the contributions of this work are detailed as follows
\begin{itemize}
    \item  We endeavor to strengthen the security of SUs operating under the threat of an eavesdropper.
    \item  The SU transmitter utilizes the EH approach to boost the energy content for transmitting messages to the destination. Additionally, we assume that the SU receiver has full-duplex (FD) capabilities and uses the PS-EH method to transmit jamming signals and deceive the eavesdropper.

    \item Assuming a CVN, we apply the cascaded Rayleigh fading model to both the main and wiretap links for realistic channel modeling.

    \item  We propose two optimization problems: one for enhancing the security of SUs by maximizing the secrecy capacity, and another for maximizing the throughput of SUs. Both problems are investigated within the constraints of energy and interference.

    \item Given the complex and dynamic nature of the CVNs environment, we suggest using a DQN technique to address the aforementioned optimization problems.
 
\end{itemize}

The remainder of the paper is as follows; the system model is described in section \rom{2}. The SUs' security optimization problem is presented in section \rom{3}. In section \rom4, the DQN methodology is outlined. The problem of SUs throughput optimization is presented in section \rom{5}. Moreover, the simulation findings are shown in section \rom{6}. Lastly, the conclusion is provided in section \rom{7}.

\IEEEpeerreviewmaketitle

 \section{System Model}

{Figure \ref{sys1}  shows several secondary users (SUs):  secondary user transmitters (SU-Tx) and   secondary user receivers (SU-Rx). SUs are communicating  over the main link $(h_s)$. In the presence of two primary users (PUs) -- a PU transmitter (PU-Tx) and a PU receiver (PU-Rx) -- the SUs are communicating using the underlay access mode. It is assumed that the SU-Tx has a limited energy supply and must harvest energy to transmit its messages. Therefore, it is presumed that the SU-Tx harvests energy from ambient sources to power the transmission of its messages. Through the wiretap channel $(h_{se})$, an eavesdropper (Eve) attempts to intercept the information shared between the SUs. The SU-Tx initially sends SU-Rx a "wake up" message. Using the technique of PS, SU-Rx extracts energy from this message. The captured energy will be stored in the battery of SU-Rx and used to impede the eavesdropper during the next symbol duration \cite{7514995}. In the next symbol duration, SU-Tx transmits the confidential information, which will be received by SU-Rx and intercepted by Eve. As a result of the FD capability of SU-Rx, the jamming signals will be transmitted utilizing the harvested energy conserved during the previous symbol duration while receiving the useful messages from SU-Tx.  Similar to \cite{7514995} and \cite{9408651}, we assume that the battery of SU-Rx charges and discharges simultaneously. In addition, it is believed that $h_s$ and $h_{se}$ are affected by numerous obstacles and objects; consequently, it is more realistic to presume that these links follow cascaded channels.  In our paper, we presume that these channels follow the cascaded Rayleigh fading model, while the remaining channels follow the single Rayleigh model, similar to \cite{9094381}. The primary objective of our work is to maximize the security and reliability level of the shared information while simultaneously enhancing the energy efficiency of the underlay CRN using the DQN strategy. It is worth mentioning that the system could  include multiple SUs that cooperate in maximizing the security, resulting in a multi-agent RL scenario.
 \begin{figure}[b]
  \centering
  \includegraphics[width=1.0\linewidth]{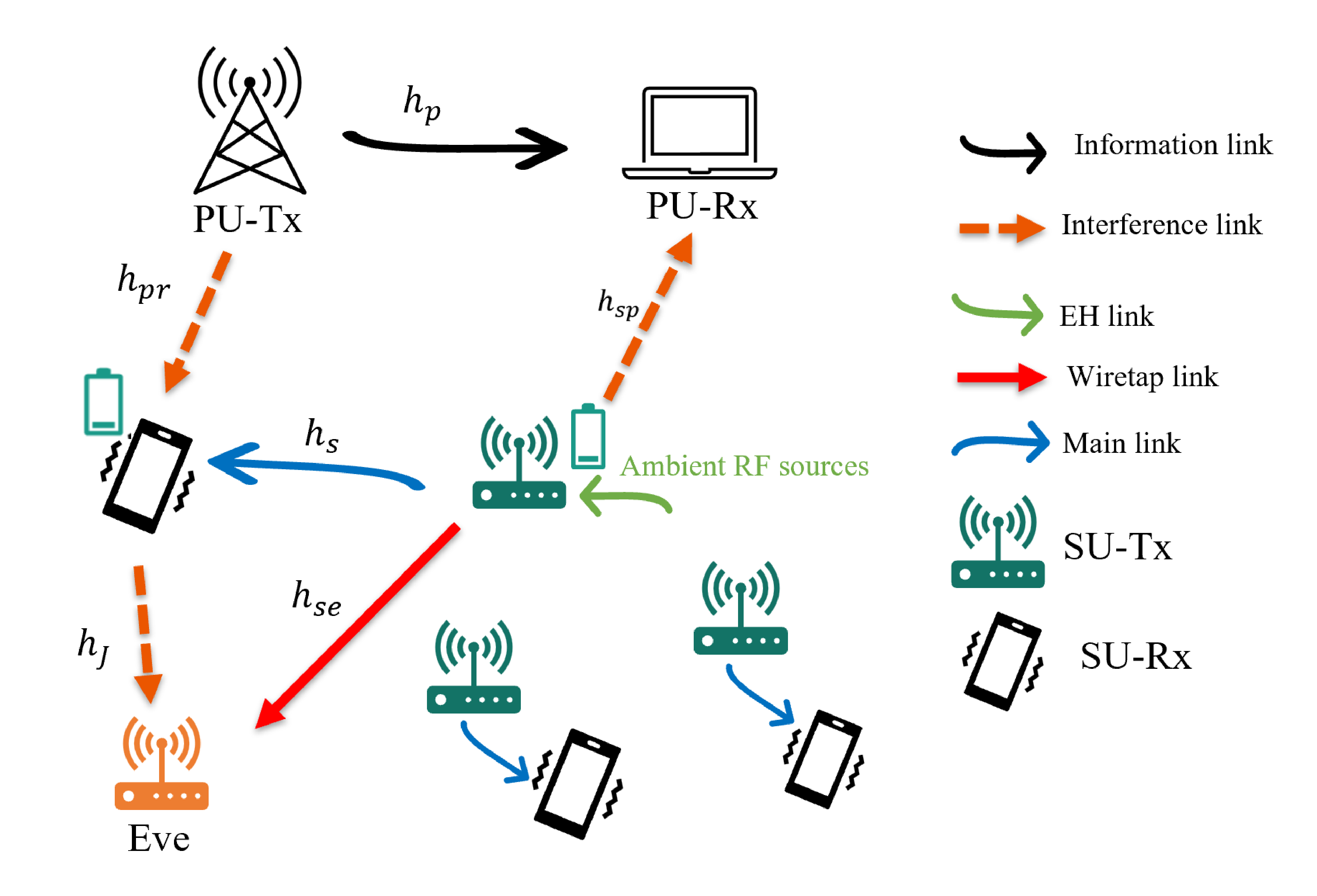}
  \caption{System model.}
  \label{sys1}
\end{figure}
\par  In this work, we assume a linear PS-EH process, which is  known for its simplicity and tractability, compared to non-linear EH approaches. Similar to  \cite{9722931} and \cite{9943999}, this assumption is valid when the average power of the energy harvested is lower than the saturation power level, as demonstrated in \cite{8435904}.   As mentioned in \cite{6740091}, the minimum acceptable received power to maintain a linearity behaviour is the EH circuits' sensitivity, which is approximately  $-27 \text{dBm}$. Nevertheless, the maximum power that a linear EH model can support is in the range of tens of milliWatts, and it is dependent upon the quality of the component, the operational frequency, the design of the rectifier circuit, and the characteristics of the diodes used  \cite{8315145}. Therefore, the linear EH model under consideration remains acceptable, due to assuming a PS-based EH, which results in a reduction in the power received at the EH receiver.  Hence, the harvested energy at SU-Rx is given by 
 \begin{eqnarray} \label{eh}
     \mathcal{E}=\theta P_s^t T_s g_s^{(p)t} \eta,
 \end{eqnarray}
 \noindent where $0<\theta<1$ is the PS factor, such that $\theta P_s^t$  is the portion employed for EH and then for jamming Eve, whereas $(1-\theta)P_s^t$ is the portion of energy   utilized by SU-Rx to decode its messages. $g_s^{(p)t}$ is the channel power gain of the EH link between the SUs $(h_{s}^{(p)})$, and it is given as $g_s^{(p)t}=|h_s^{(p)t}|^2$. $\eta$ represents the energy conversion efficiency coefficient and  $T_s$ is the transmission time slot. 
 Given  (\ref{eh}), the jamming power transmitted by SU-Rx  is given by
   \begin{IEEEeqnarray}{lCr}  \label{pj}
P_j= \theta P_s^t g_s^{(p)t} \eta.
\end{IEEEeqnarray}

The PU-Tx is assumed to operate in a slotted mode, i.e., for a total of $N$ slots, PU-Tx is active for the first $\xi$ slots, and the band is considered inactive for the remaining $N-\xi$ slots.  It is assumed that the SU-Tx identifies when PU-Tx is active, which can be determined by analyzing the intensity of the received PU's signals using a variety of approaches, such as the energy detection technique \cite{9237455},\cite{10182973}. However, the SU-Tx will respect the interference threshold tolerable at the PU-Rx during all time slots. Moreover, using the transmitter verification technique \cite{4286333}, the SU-Tx can determine the location of the PU transmitter by analyzing the signal's characteristics. SU-Tx also sets an indicator for the PU's status \cite{9645987}, represented by $\mathcal{D}_t$ as
\begin{IEEEeqnarray}{lCr} 
 \mathcal{D}_t= \begin{cases}
    1, & \text{channel is occupied by PUs in the $t^{th}$ slot} \\
    0 , & \text{channel is idle in the $t^{th}$ slot } 
 \end{cases} \cdot
\end{IEEEeqnarray}
\noindent We further assume that the PUs' transmissions have a negative effect on the communications of the SUs during the initial $\xi$ time intervals, and are therefore considered interference to their communications.
Hence, the received message at the SU-Rx is given by
\begin{eqnarray}\label{ys}
    y_s=\sqrt{(1-\theta)P_s^t} h_s^t x_s+ \mathcal{D}_t \sqrt{P_p^t} h_{pr}^t x_p + n_s,
\end{eqnarray}
\noindent  where  $x_s$ is the transmitted SUs' messages, and $n_s$ is the additive-white-Gaussian-noise (AWGN) at SU-Rx with zero mean and variance $N_0$.  $P_p^t$ is the transmission power of PU-Tx, $h_{pr}^t$ is the channel gain between PU-Tx and SU-Rx, and $x_p$ is the transmitted PUs' message symbol.   Without compromising generality, in (\ref{ys}), we presume that the destination is able to cancel the self-interference caused by the FD property using one of the self-interference cancellation techniques \cite{6517516,5089955}. Moreover, the intercepted message by the eavesdropper is given as
\begin{eqnarray}\label{ye}
    y_e=\sqrt{P_s^t} h_{se}^t x_s+\sqrt{P_{j}^t} h_j^t x_j+n_e,
\end{eqnarray}
\noindent where $h_{se}^t$ is the wiretap channel gain, $P_j^t$ is the jamming power delivered by SU-Rx, $x_j$ is the jamming signal, and $n_e $ is the  AWGN  at Eve with zero mean and variance $N_0$.

At each time period, SU-Tx decides whether to broadcast its messages or harvest energy. Consider $\mu_t$ to be an index for the selected option at each time slot, with two possible values given as
\begin{IEEEeqnarray}{lCr} 
\mu_t= \begin{cases}
    1, & \text{harvest energy during the $t^{th}$ slot} \\
    0 , & \text{transmit messages during the $t^{th}$ slot} 
 \end{cases} .
\end{IEEEeqnarray}

To prevent interfering with the PU-Rx, it is crucial to ensure that the power transmitted by the SU-TX at each time slot $(P_s^t)$ does not exceed the   interference threshold tolerable at PU-Rx.  As a result, the interference restriction is expressed as \cite{9348134}
\begin{IEEEeqnarray}{lCr} 
 \mathcal{D}_t g_{sp}^t P_s^t \leq I_{th}^t,
\end{IEEEeqnarray}
\noindent where $I_{th}^t$ is the interference threshold tolerable at  PU-Rx   and $g_{sp}^t$  is the channel power gain between SU-Tx and PU-Rx given as $g_{sp}^t=\left|h_{sp}^t\right|^2$, with $h_{sp}^t$ is the channel gain  between SU-Tx and PU-Rx. 

  %SNRs:
  For the first $\xi$ time slots, PU-Tx is assumed to be active, and thus it will interfere with the reception of the SU-Rx. Hence, given (\ref{pj}) to (\ref{ys}), the signal-to-interference-plus-noise ratio (SINR) at SU-Rx is given by
  \begin{eqnarray}\label{snrsa}
      \gamma_s^A=\frac{(1-\theta) P_s^t g_s^t}{P_p^t g_{pr}^t +N_0}.
  \end{eqnarray}
\noindent During the remaining time slots, i.e., $N-\xi$, PU-Tx is idle and thus the signal-to-noise ratio (SNR) at SU-Rx is given by
  \begin{eqnarray}\label{snrsi}
      \gamma_s^I=\frac{(1-\theta) P_s^t g_s^t}{N_0}.
  \end{eqnarray}
  \noindent Moreover, the SINR at Eve is given by 
    \begin{eqnarray}
      \gamma_e=\frac{P_s^t g_{se}^t } {P_j^t g_{j}^t +N_0},
  \end{eqnarray}
 \noindent where $g_j^t$ is the channel power gain of the jamming link, such that $g_j^t=|h_j^t|^2$. 
 \iffalse
 Accordingly,  the data rate achieved at SU-Rx and Eve are given, respectively, as 
 \begin{eqnarray} \label{rate}
 R_s=\log_2 (1+\gamma_s^i), 
 \end{eqnarray}
 \begin{eqnarray} \label{ratesu}
 R_e=  \log_2 (1+\gamma_e),
 \end{eqnarray}
\noindent for $i\in \{A,I\}$. \fi

We assume that the main and wiretap links follow the cascaded Rayleigh fading model.  Therefore, let  $h_{i}^t=\prod_{j=1}^{N_u} x_j$, for $i\in\{s, se\}$ and $u \in\{m,e\}$.  $x_j$  follows the Rayleigh model, and  $N_u$ represents the cascade level of the  link  $h_{i}$, such that $N_m$ is the main link $(h_s)$ cascade level and $N_e$ is the wiretap link $(h_{se})$ cascade level \cite{9348134}. The probability density function (PDF) of $h_{i}^t$ is evaluated using the transformed  Nakagami-$m$ distribution   \cite{lu2011accurate}. Hence, the  PDF  of the channels power gain    $(g_{i}^t=\left|h_i^t\right|^2)$  is given as      
\begin{eqnarray} \label{pdfcasc}
f_{g_{i}^t} (y)&=&   \frac{\beta_i}{2} y^{\frac{m_i}{N_u}-1} \exp\left(-\frac{m_i}{\Omega_i \sigma_i^{\frac{2}{N_u}}} y^{\frac{1}{N_u}}\right),
\end{eqnarray}
\noindent where    $\beta_i=\frac{2 \left(\frac{m_i}{\Omega_i}\right)^{m_i} }{N_u \Gamma (m_i) \sigma_i^{\frac{2{m_{i}}}{N_u}}}$, $\sigma_i$ is the scale parameter,     and $\sigma_i^2=\prod_{j=1}^{N_{u}} \sigma_{i_{j}}^{2}$. $m_i$ and $\Omega_i$ are   calculated as 
 \begin{eqnarray}  
m_i&=& 0.6102 N_u+0.4263, \;\;\;   \nonumber \\ \Omega_i &=&0.8808 N_u^{-0.9661}+1.12.
\end{eqnarray}
\noindent  Moreover, the PDF   of the rest of the links, i.e.,  $h_{w}^t$, for $w \in \{p,sp,pr,j\}$  follows  the single Rayleigh model, and thereby the PDF and the cumulative distribution function (CDF) of the channel power   gain $(g_{w}^t)$ follow  the exponential distribution, respectively, as 
\begin{eqnarray} \label{hre}
f_{g_{w}^t} (y)&=&   \lambda_{w}  \exp\left(-\lambda_{w} y\right),
\end{eqnarray}
\begin{IEEEeqnarray}{lcr} \label{cdfre} 
F_{g_{w}}(y)=1-\exp{\left(-\lambda_{w} y\right) },
\end{IEEEeqnarray}
\noindent with $\lambda_{w}$ being the fading channel parameter. 
 
We presume that the highest capacity of the battery at SU-Tx is denoted by $C_{max}$, while the initially available energy is represented by $C_i$. Since the SU-Tx employs the stored energy to power transmissions, the battery's energy content will fluctuate based on whether the SU-Tx is harvesting energy or transmitting messages. If $\mu_t$ is equal to one, for instance, SU-Tx will collect and store $E_{h}^t$ of energy in the battery. Similar to \cite{9645987}, the process of harvesting from ambient RF sources is to be modeled at each time interval as a uniform distribution extending from 0 to $E_{max}$.  If SU-Tx decides to transmit messages, the amount of available energy in the next time period $(C_{t+1})$ can be altered as follows  \cite{9448276}
\begin{IEEEeqnarray}{lCr} \label{ensure}
  C_{t+1}= \min\{C_t+\mu_t E_h^t-\left(1-\mu_t\right)   P_s^t T_s, C_{max}\}, 
\end{IEEEeqnarray}
\noindent where $C_t$ denotes the available energy at the beginning of the $t^{th}$ slot and   $\left(1-\mu_t\right)   P_s^t {\color{black}T_s}$ represents the utilized energy for transmission.  
Notably, (\ref{ensure}) ensures that the stored energy does not exceed the maximum capacity of the SU-Tx battery. 

 \section{Secondary Users' Security Optimization}
This scenario's primary objective is to maximize the SU's secrecy rate by rendering it resistant to eavesdropping. Hence, the problem can  be formulated as  
  \begin{align} \label{opti-prob1}
   \mathcal{P}1: \;\;  & \underset{\mu_t,P_s^t}{\text{max}}
    & &  Sec_{c}^i \\
    & \text{s.t.}  
    && \label{firstrate1} (1-\mu_t)P_s^t T_s\leq C_t, &  \\ &&& \label{lastcon11}  \mathcal{D}_t g_{sp}^t P_s^t \leq I_{th}^t, 
    &  
  \end{align}
\noindent where $Sec_{c}^i$, for $i\in\{A,I\}$, is the secrecy capacity,  which is  defined as the maximum rate at which confidential information can be transmitted over a communication channel with the presence of eavesdroppers. $Sec_{c}^i$ is  given by
\begin{eqnarray} \label{cap-eq} 
Sec_{c}^i= \max \{M_{c}^i-E_{c},0\} ,
\end{eqnarray} 
\noindent  with   $M_{c}^i=  \log_2(1+\gamma_s^i)$ is the capacity of the main link $(h_{s})$ and  $E_{c}= \log_2(1+\gamma_{e})$ is the wiretap link $(h_{se})$ capacity. $\mathcal{P}1$ ensures maximizing the secrecy rate of the SUs while guaranteeing that the selected energy for transmission does not exceed the current energy content of the battery, as demonstrated by constraint (\ref{firstrate1}). In addition, constraint (\ref{lastcon11}) ensures compliance with the underlay access mode interference threshold. $Sec_{c}^A$ is the secrecy capacity when PU-Tx is active and it is given by
\begin{eqnarray}\label{capa}
    Sec_{c}^A&=&M_c^A-E_c \nonumber \\ &=& \log_2\left(\frac{1+\gamma_s^A}{1+\gamma_e}\right),
\end{eqnarray}
whereas $Sec_{c}^I$ is the secrecy capacity when the PU-Tx is idle and it is similarly given as
\begin{eqnarray}\label{capi}
   Sec_{c}^I&=&M_c^I-E_c \nonumber \\ &=& \log_2\left(\frac{1+\gamma_s^I}{1+\gamma_e}\right),
\end{eqnarray}
\noindent where $\gamma_s^A$ and $\gamma_s^I$ are  defined in   (\ref{snrsa})   and (\ref{snrsi}), respectively.
\section{DQN Approach to Optimize $P_s^t$}
%MDP:
The problem described by (\ref{opti-prob1}) can be modeled as a model-free Markov decision process (MDP). This is because in each slot, the state of the current slot depends solely on the state of the previous slot, thus satisfying the Markov property.  Hence, the model is comprised of the four tuples $<S, A, R, T>$. $S$ represents the states space, and thus the state at each time slot $(S_t)$ contains the following combination
\begin{IEEEeqnarray}{lcr}\small
    S_t=\{\mathcal{D}_t, E_{h_{t-1}},C_t, g_{s^{(p)}}^t, g_{s}^t, g_{pr}^t,  g_{sp}^t,g_{p}^t, g_{j}^t, g_{se}^t\}, 
\end{IEEEeqnarray}
\noindent \noindent where $E_{h_{t-1}}$ represents the harvested energy during the previous time slot. Moreover, the action space $A$ represents the actions that could be performed by the agent. The action at each time interval specified by $A_t$ is expressed as
\begin{IEEEeqnarray}{lcr}
   A_t=\{\mu_t, P_s^t\}.
\end{IEEEeqnarray}
\noindent Thus, SU-Tx must determine whether to collect energy or send messages, as well as the transmit power $(P_s^t)$ value for each time slot.  In addition, each time slot's reward function is provided by
\begin{IEEEeqnarray}{lCr} \label{reward1}
\small
R_t= \begin{cases}
    Sec_{c}^A, & \mu_t=0, \mathcal{D}_t=1, P_s^t g_{sp}^t \leq I_{th}^t  \\
    Sec_{c}^I , & \mu_t=0, \mathcal{D}_t=0, P_s^t g_{sp}^t \leq I_{th}^t  \\
    0 , & \mu_t=1,   P_s^t T_s > C_t \\
    -\zeta , & \text{else}
 \end{cases} .
\end{IEEEeqnarray}
%Third line means, if available energy (C_t) was low, i.e., C_t was lower that need transmission energy, then it will not go to the first two rows of the reward, it will harvest energy instead (\kappa_t=1). 
\noindent It is evident from (\ref{reward1}) that there are multiple options for the agent's imminent reward. The reward may be dependent on the possible secrecy rates, i.e. (\ref{capa}) and (\ref{capi}). Additionally, the agent receives no reward (data rate) if the SU-Tx decides to harvest and the selected energy for transmission exceeds the current energy content of the SU's battery. Moreover, if the agent fails to comply with any of the specified requirements, it is penalized for an amount of $-\zeta$, where $\zeta$ is a positive number.  Finally, a time step $T$ consists of time periods. A step is the transition from state $s_t$ to the next state $(s_{t+1})$. The state-action pairing is repeated until all available time segments expire. As our work involves environments with discrete state and action spaces, we opted to implement the DQN method on account of its applicability in such environments. That is, it is well-known that the DQN method operates efficiently and discovers optimal policies in discrete action and state spaces. Moreover, since CRNs are renowned for their extreme complexity and dynamism, the DQN approach represents a viable solution as the agent in DQN gradually acquires knowledge and modifies its policies,  exhibiting a degree of adaptability.
  
The primary objective of our DQN strategy is to train the agent, which in our case is the SU-Tx,  to interact with the environment to maximize the future accumulated reward \cite{9351818,9729992,10592377}. In other words, the continuous interaction between the agent and the environment will maximize the SUs' long-term security. In a given environment, a deep neural network will estimate Q-values for each state-action combination and get close to its optimal Q-function \cite{9524882}. It develops the ability to map state-action pairings to Q-values that indicate the anticipated reward when performing a certain action in a specific state, taking into consideration the potential reward for adhering to a specific policy.  In DQN, the Q-function is given by \cite{9645987}
\begin{IEEEeqnarray}{lCr}  \label{Q-ini}
  Q^{\pi}\left(s,a\right)= E_{s^{\prime}, a^{\prime}}\left( r+\nu Q^{\pi} \left(s^{\prime},a^{\prime}\right)|s,a\right),
\end{IEEEeqnarray}
\noindent where $s$ represents the current state and $a$ represents the chosen action. $a^\prime$ is the action chosen at state $s^\prime$, which is the state reached after the action $a$ is performed. $r$ represents the immediate reward for selecting action $a$. $E[\cdot]$ represents the expectation operator. The discount factor  $\nu$ is a scalar value between 0 and 1, which determines the value of future rewards. It specifies how much consideration should be given to future rewards when making decisions.
In addition, $\nu$ is utilized to balance the trade-off between immediate and future rewards. $\pi$ represents the agent's policy for selecting actions in a given state. Due to the extremely dynamic environment, it is important to note that the $\epsilon$-greedy strategy is used to determine the optimal transmit power of the SU-Tx. During training, the algorithm employs an $\epsilon$-greedy exploration policy, where the agent selects a random action with probability $\epsilon$ and the action that maximizes the Q-value for the current state with probability $1-\epsilon$.
 The utilized decay equation for this policy is given as \cite{mnih2015human}
\begin{eqnarray}   \epsilon=\epsilon_{min}+\left(\epsilon_{max}-\epsilon_{min}\right) e^{-r_d t},
\end{eqnarray}
\noindent where $\epsilon_{min}$ represents the minimum value of $\epsilon$, $\epsilon_{max}$ represents the maximum value of $\epsilon$ (typically set to 1), and $r_d$ represents the decay rate. To maximize the cumulative long-term return, we must identify the optimal action in each time period that maximizes the state-action value defined in (\ref{Q-ini}) as  \begin{IEEEeqnarray}{lCr}  
  a^\ast =\underset{a} \arg \max  Q^{\pi}\left(s,a\right) .
\end{IEEEeqnarray}
\noindent In DQN approach, neural networks are used to approximate the Q-value function by updating the neural network's weights and biases, represented by the parameter $w_n$. 
The approximation is given as 
\begin{IEEEeqnarray}{lCr}  
   Q\left(s,a;w_n \right)\approx  Q^{\ast}\left(s,a\right) ,
\end{IEEEeqnarray}
\noindent where $ Q^{\ast}\left(s,a\right)=E_{s^{\prime}} \left(r+\nu \max_{a^{\prime}} Q^{\ast}\left(s^{\prime},a^{\prime}|s,a\right)\right)$ being the optimal value function. Using the loss between  the predicted    and the target Q-values, the parameter $w_n$ is updated via gradient descent and backpropagation. The loss function $(\mathcal{L})$ is given by the mean square error as \cite{9840876}
\small
\begin{IEEEeqnarray}{lCr}  \label{loss}
  \mathcal{L}(w_n)=E\left(\left( r_s+\nu \max_{a^{\prime}} Q^{\pi} \left(s^{\prime},a^{\prime};w_n \right)-Q^{\pi}\left(s,a;w_n\right)
  \right)^2\right)   , \nonumber \\
\end{IEEEeqnarray}
\normalsize
\noindent where the first portion of  (\ref{loss})  $\left( r_s+\nu \max_{a^{\prime}} Q^{\pi} \left(s^{\prime},a^{\prime};w_n \right)\right)$  indicates the desired Q-value, and the second component is the predicted Q-value, which is periodically updated during the learning process. $r_s$  represents the reward received for moving from state $s$ to state $s^\prime$.
The stochastic gradient descent (SGD) algorithm can be used to minimize the loss function by updating $w_n$ in the direction that minimizes the loss. The SGD rule is expressed as
\begin{IEEEeqnarray}{lCr}  \label{sgd}
y=\alpha \frac{\partial  \mathcal{L}(w_n) }{\partial w_n}    ,
\end{IEEEeqnarray}
\noindent where $0<\alpha<1$ represents the learning rate, which determines how rapidly the Q-value function changes depending on the measured rewards and the present Q-value function estimates.  Lastly, using (\ref{sgd}), $w_n$ is updated as follows 
\begin{IEEEeqnarray}{lCr}   
w_n=w_n-y .
\end{IEEEeqnarray}
 
\section{Secondary Users' Throughput Optimization}  

There are several reasons why optimizing the throughput of SUs in an underlay CRN is crucial.  First, the system's overall efficiency and capacity are enhanced, allowing for more simultaneous data transmission and reception. This boosts network performance and user satisfaction. Second, in an underlay cognitive radio system where SUs and PUs simultaneously access the spectrum, optimizing the throughput ensures that SUs can utilize the available spectrum resources efficiently without interfering with PUs. Third, the fact that the SU transmitter harvests energy from ambient RF sources intensifies the importance of throughput optimization. Ambient EH enables SUs to extend their operational lifetime and achieve self-sufficiency, thereby decreasing their reliance on conventional power sources. By optimizing throughput, we can increase the energy efficiency of SUs, allowing them to transmit and receive data for a longer duration using the harvested energy. This increases the network's total operational duration and enables SUs to provide uninterrupted communication services.
Therefore, in this section, the objective is  to maximize the throughput of the SUs $(M_{c}^i)$ while ensuring that their transmission power does not exceed the permissible interference threshold level at PU-Rx.  Consequently, the problem can be stated as follows
 \begin{align} \label{opti-prob}
   \mathcal{P}2: \;\;  & \underset{\mu_t,P_s^t}{\text{max}}
    & & M_{c}^i \\
   & \text{s.t.}  
    && \label{firstrate} (1-\mu_t)P_s^t T_s\leq C_t, &  \\ &&& \label{lastcon1}  \mathcal{D}_t g_{sp}^t P_s^t \leq I_{th}^t, 
    &  
  \end{align}

\noindent To solve this problem, we use the same  DQN approach utilized in section \rom{4}. However, the agent immediate reward function is updated as
\begin{IEEEeqnarray}{lCr} \label{reward}
\small
R_t= \begin{cases}\label{rew2}
    M_c^A, & \mu_t=0, \mathcal{D}_t=1, P_s^t g_{sp}^t \leq I_{th}^t  \\
    M_c^I , & \mu_t=0, \mathcal{D}_t=0, P_s^t g_{sp}^t \leq I_{th}^t  \\
   0 , & \mu_t=1,   P_s^t T_s > C_t \\
    -\zeta , & \text{else}
 \end{cases} .
\end{IEEEeqnarray}

\section{Simulation results}
In this section, the results for the proposed system model and analyses are presented.   It is worth mentioning that the selection of the parameters' value is influenced by the ones utilized in \cite{10182973} and \cite{9645987}. To consider the path loss   effect on the system performance, we   assume that the SU transmitter  represents the reference location in a two-dimensional space. That is, SU-Tx  is located at $(0,0)$ and the other nodes (SU-Rx, Eve, PU-Tx, and PU-Rx) are of different distances from  SU-Tx. Assume $ d_{y}^{-PL}=\frac{1}{2 \lambda_J}$, such that $\lambda_J=\frac{1}{2\sigma^2}$, where $d_y$ represents the distance between the nodes connected by the link $y$, for $y\in\{p, sp, pr, j, s, se \}$, in meters $(m)$. $PL$ is the path loss exponent.  Through the entire section, we assume that $P_{p}^t$ is varying randomly in the range from $0$ to $1$ Watt $(W)$. Moreover, unless otherwise mentioned in the figures' caption, we assume that   $N=20$,   $\nu=0.99$, $\alpha=0.003$,    $\zeta=1$,   $PL=2$.  $T_s=1$ second, $N_0=1$,  $E_{max}=0.2$ Joule $(J)$,  $C_{max}=0.5 J$,  $\epsilon_{max}=1$, $\epsilon_{min}=0.01$, and  
 $r_d=0.001$. The distances between nodes are given as  $d_{sp}=500 m$, $d_{se}=100 m$  , $d_{pr}=500 m$, and $d_j=50 m$.

Figure \ref{one} depicts the average SU reward, which represents the average secrecy rate of SUs. Using the suggested DQN method, convergence is observed, indicating that the DQN method optimizes long-term performance. In addition, as the channel between SU-Tx and SU-Rx (main link) cascade level $(N_m)$ increases, indicating an increase in the number of objects and obstacles, a decline in the average reward is observed. This is due to the fact that a greater $N_m$ results in a more severe fading on the main link, rendering a reduced link's reliability and, consequently, a lower secrecy rate and reward.

%Cascaded Section
 \begin{figure}  [t!] 
  \centering
  \includegraphics[width=0.9\linewidth]{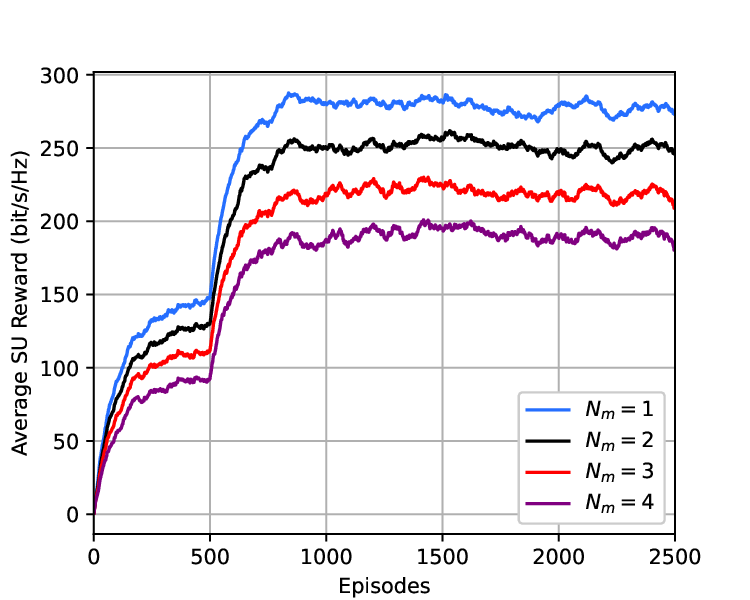}
  \caption{Average SUs' reward. $N_e=2$, $\eta=0.9$, $\theta=0.9$, $d_s=2 m$,  $\xi=2$, and $I_{th}=0.01 W$. }
  \label{one}
\end{figure}

Figure \ref{two} depicts the average SUs' reward in relation to the PS factor at the SU-Rx $(\theta)$ for various values of the EH efficiency $(\eta)$. The average secrecy rate can be noticed to be a concave function of $\theta$. Specifically, as $\theta$ increases, indicating a greater quantity of energy is harvested, the level of security improves. This is because this harnessed energy is used to create interference signals to confuse the eavesdropper, thereby enhancing the security of the shared communications. After a certain value of $\theta$, however, the security begins to decline. This is because the quantity of received energy utilized to decode the useful messages at SU-Rx will decrease, resulting in a reduction in the quality of the decoded messages and a decline in security.  In light of this, it is imperative to select the value of $\theta$ carefully in order to enhance security, given that the energy extracted has an effect on reliability of the system and, consequently, security. In addition, as $\eta$ increases, the harvesting efficiency improves, which increases the quantity of energy harvested for the purpose of obstructing the eavesdropper, thereby enhancing security.
 
  \begin{figure}  [b!] 
  \centering
  \includegraphics[width=0.9\linewidth]{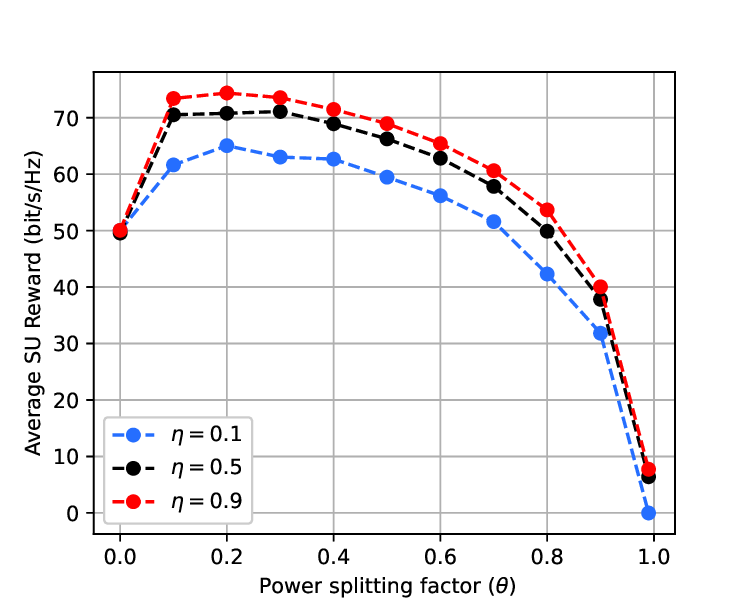}
  \caption{ Average SUs' reward versus the power splitting factor $(\theta)$.     $N_e=2$, $N_m=1$,  $d_s=12 m$, $\xi=6$, and $I_{th}=0.01 W$.}
  \label{two}
\end{figure}

Figure \ref{three} depicts the average secrecy rate of SUs for various values of  active time periods for the PU-Tx $(\xi)$.   One can conclude that the average SU reward decreases as $\xi$ increases. In our work, we presume that SU-Rx is near to PU-Tx, and this proximity can be adversely impacted by the PUs' communication. Therefore, the longer the PU is active, the less secure the SUs' transmissions. However, this effect can be mitigated by moving the SU-Rx closer to the SU-Tx, as channel reliability improves when distance decreases. This is demonstrated by reducing the distance $d_s$ from $15 m$ to $5 m$ for $\xi=10$. Moreover, it is evident that the curve associated with $\xi=N$ yields an unsatisfactory value of the secrecy rate as it descends below zero.

   \begin{figure}  [t!] 
  \centering
  \includegraphics[width=0.9\linewidth]{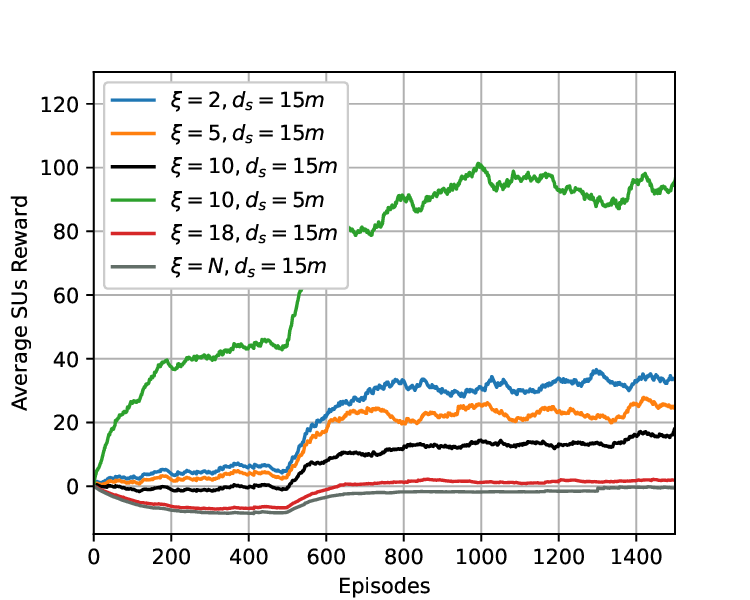}
  \caption{ Average SUs' reward for    $N_e=3$, $N_m=4$,     $\theta=0.4$, $\eta=0.1$, and $I_{th}=0.01 W$.}
  \label{three}
\end{figure}

 The impact of the interference threshold tolerable at PU-Rx $(I_{th})$ is shown in Figure \ref{three_2} for two values of the maximum battery capacity of SU-Tx $(C_{max})$. As $I_{th}$ increases, the communication between SUs becomes more confidential. This is due to the fact that as $I_{th}$ rises, SU-Tx is permitted to increase its transmission power, which improves the received SNR at SU-Rx, thereby enhancing security. 
 When the interference threshold is very low, the SU transmitter's transmission decisions must be performed with caution to avoid exceeding the threshold. As the SU transmitter's maximal battery capacity increases, it can harvest and store more energy. As a result of being penalized for exceeding the underlay threshold, the DQN agent may initially prioritize energy harvesting over transmission in order to reduce the risk of interference and, consequently, the  security degrades  as $C_{max}$ increases.   As $I_{th}$ rises, however, the DQN agent modifies its decision-making policy and begins to prioritize transmission over energy harvesting without concern of violating the threshold constraint. This modification enables the SU transmitter to effectively utilize the stored energy and increase its secrecy rate. Therefore, the observed behavior changes when the DQN agent recognizes the increased transmission capacity and modifies its strategy accordingly. Moreover,  Figure \ref{Ith_accept} demonstrates which values of $I_{th}$ result in invalid secrecy rates, namely those that are equal to or less than zero. It is observed that the agent  begins to receive unacceptable secrecy rate values for $I_{th}<-77 dB$.

 %Furthermore, results indicate that the average secrecy rate of SUs improves as $C_{max}$ becomes higher. This is because a larger battery capacity indicates that the SU-Tx has the capability to capture more energy, resulting in a greater level of transmission power. Consequently, this will increase the SNR at SU-Rx and thereby the average secrecy rate of the SUs' transmissions. 

   \begin{figure}  [b!] 
  \centering
  \includegraphics[width=0.9\linewidth]{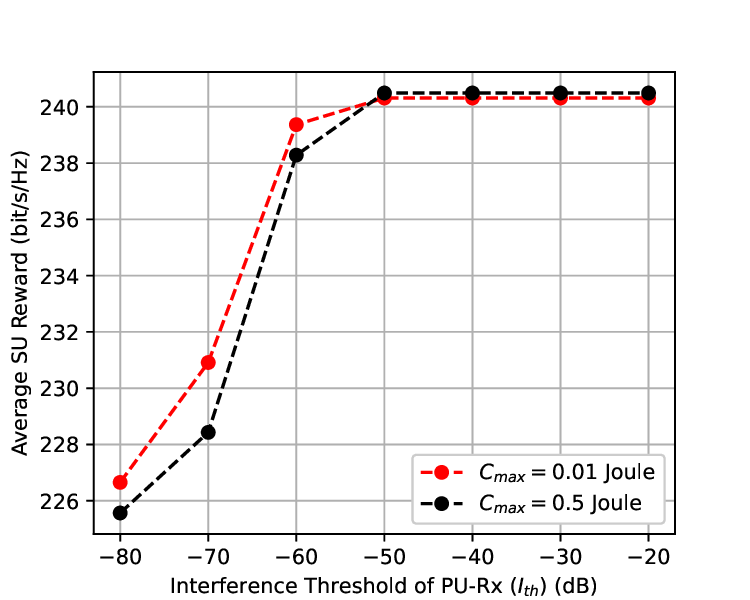}
  \caption{Average SUs' reward versus $I_{th}$.     $N_e=1$, $N_m=1$,  $d_s=15 m$,   $\xi=2$, $\theta=0.6$, and $\eta=0.9$.}
  \label{three_2}
\end{figure}

   \begin{figure}  [t!] 
  \centering
\includegraphics[width=0.9\linewidth]{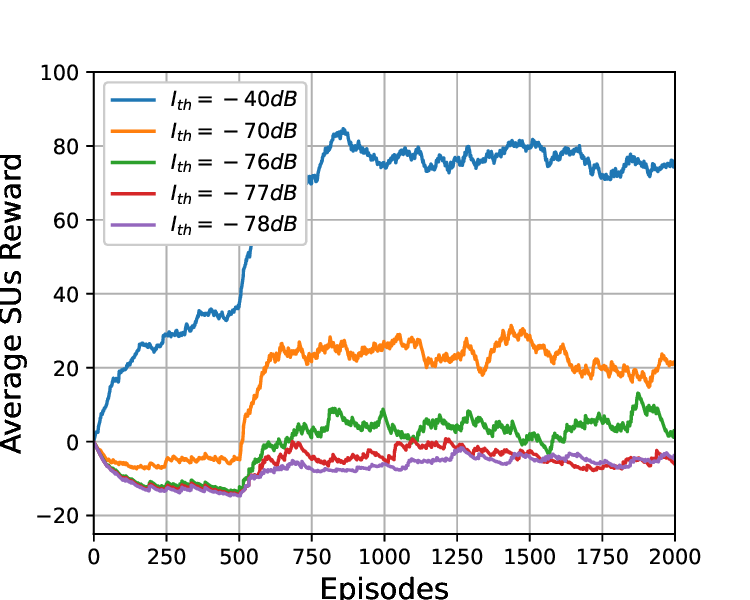}
  \caption{ Average SUs' reward for   $N_e=1$, $N_m=1$,  $d_s=15 m$,   $\xi=2$, $\theta=0.4$, and $\eta=0.9$.}
  \label{Ith_accept}
\end{figure}

A comparison between the proposed DQN approach, a \textit{random policy}, and a \textit{harvest/transmit} policy  is illustrated in Figure \ref{three_3}. A \textit{random policy} represents a situation in which the agent selects an action at random with no learning from its surroundings. Specifically, the agent will choose a random transmission power value and will randomly choose whether to broadcast or harvest throughout each time interval. In the first stages of the process,  we found that the outcomes of the DQN and the random policies are consistent with one another. This is because  during the training phase, the agent using the DQN method starts investigating its surroundings rather than taking advantage of its most advantageous possibilities. This is due to the agent starting its exploration with $\epsilon=1$, as required by the $\epsilon$-greedy policy. This implies that the agent will behave at random with a probability of \textit{one}, allowing it to completely investigate its surroundings.  However, the value of $\epsilon$ continues to fall as the agent becomes more familiar with the environment and learns which actions are more likely to lead to significant rewards.  After the agent determines the appropriate course of action, it will begin to exploit those choices, which will provide the largest future return compared to the random method. 
 Additionally, a \textit{harvest/transmit} policy specifies the scenario where the agent only harvests energy  during the $\xi$ time slots. During the remaining  time slots, the SU-Tx only transmits messages and randomly determines the transmission power at each time slot.  The results indicate that the DQN method exhibits superior performance compared to the harvest/transmit method. The harvest/transmit method exclusively transmits during the $N-\xi$ time intervals, failing to account for energy lost during transmission. This deficiency in reliability compromises the main link's security. Furthermore, SU-Tx solely collects energy without transmitting messages during the initial $\xi$ time intervals, thereby diminishing the rate and reliability of the SUs' link.

    \begin{figure}  [b!] 
  \centering
  \includegraphics[width=0.9\linewidth]{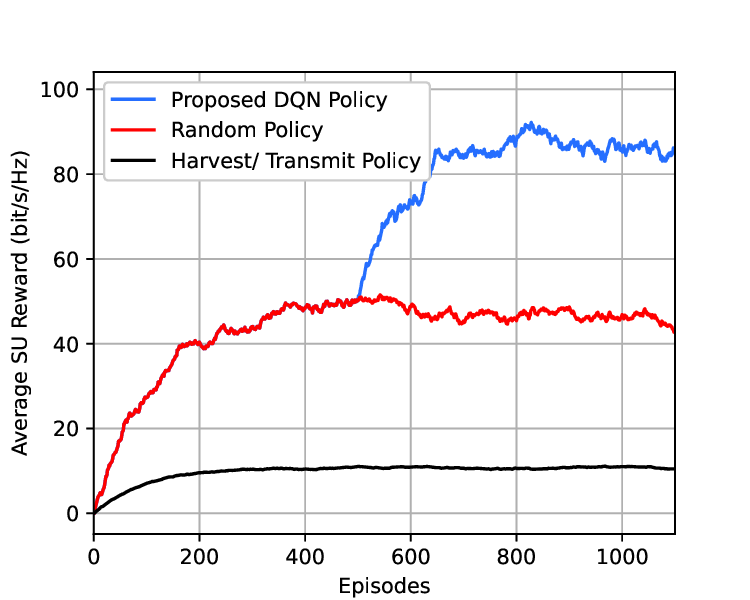}
  \caption{ Average SUs' reward.     $N_e=2$, $N_m=2$,  $d_s=15 m$,   $\xi=2$, $\theta=0.8$, $\eta=0.9$, and $I_{th}=0.01 W$.}
  \label{three_3}
\end{figure}

    \begin{figure}  [h!] 
  \centering
  \includegraphics[width=0.9\linewidth]{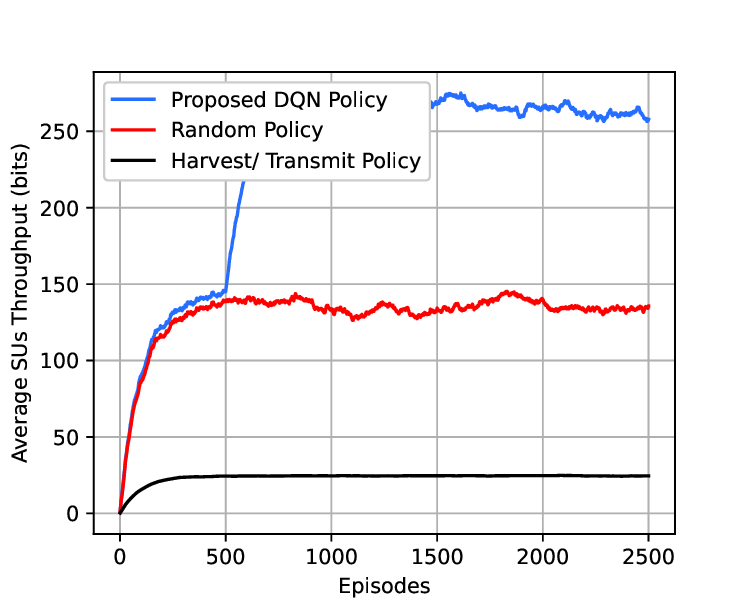}
  \caption{ Average SUs' throughput.     $N_e=2$, $N_m=1$,  $d_s=15 m$,   $\xi=2$, $\theta=0.6$, $\eta=0.9$,  and $I_{th}=0.01 W$.}
  \label{four}
\end{figure}

  Figure \ref{four} presents the average SUs' throughput problem presented in $\mathcal{P}2$. We also compare the average SUs' throughput for the DQN approach, random policy, and harvest/transmit policy. The DQN strategy, in which the agent maximizes the rate of the SUs by intelligently selecting transmission power and determining whether to harvest or transmit in each time period, outperformed the other two strategies. This is because the DQN algorithm is capable of learning and optimizing its decision-making process based on environmental feedback. By continuously modifying its policy using RL, the DQN agent is able to adapt and make more informed decisions, resulting in a higher throughput of SUs. Moreover, the random policy performed better than the harvest/transmit scheme. Although the random policy lacks intelligence and systematic decision-making, its random nature can occasionally result in more favorable outcomes and higher performance than a fixed scheme such as harvest/transmit. However, it should be noted that, unlike the DQN approach, the random policy is not designed to learn or progress over time.  The harvest/transmit scheme exhibited the lowest average throughput of SUs due to the scheme's lack of intelligent decision-making and optimization.   Furthermore, the limited number of accessible time periods, namely $N-\xi$, for potential transmission significantly decreases the probability of transmission and hence reduces the throughput of the SUs.

%%--------------------------------------------

\section{Conclusions}
This paper investigates the application of RL to maximize the PLS of an underlay CRN when an eavesdropper is present. The SU transmitter harvests energy to fuel its transmissions, while the SU receiver harvests energy to send jamming signals to limit the eavesdropper's intercept capabilities.  The SU transmitter serves as a DQN agent that must maximize the security and reliability of its transmissions by determining the transmit power and whether to transmit or harvest energy at each time interval. Our findings indicate that the DQN approach converges after a specified number of episodes and that, given the interference constraints of primary users and limited battery capacity, the achievable long-term secrecy rate and throughput are maximized. In addition, the cascade level  has a significant impact on the security level, indicating that it should not be disregarded in such circumstances. To maximize the security of the SUs' messages, our findings also suggest that the SU receiver should carefully select the power splitting factor that dictates the amount of harvesting versus the amount used for decoding. In addition, security appears to improve as the  interference threshold of the PUs increases. In terms of security and reliability, our findings demonstrate that the utilized DQN approach outperforms the random policy and the fixed harvest/transmit policy.

%This paper investigates an overlay cognitive radio network with two primary users (PUs) and several secondary users (SUs). In exchange for accessing the licensed band, one of these multiple SUs is selected to forward the PUs messages using the harvested energy  from the PU transmitter messages. Our results indicate that a higher density of these SUs is required to increase the link reliability of the PUs and SUs communication.  The outage probability of both links and their asymptotic expressions have been derived.   In addition, the time switching factor and the SU relay power allocation factor are optimized for two scenarios: maximizing the SUs' rate while constraining the PUs' rate, and maximizing the sum of both networks' rates. Our results demonstrate that, as compared to  fixed factors, the derived optimized ones achieve the optimum performance.  


\end{document}